\documentclass{PoS}

\usepackage{amsmath, amssymb,graphics,epstopdf,color,cancel,graphicx}
\usepackage[utf8]{inputenc}

\usepackage{lineno}

\title{Constraints on cross-section and lifetime of dark matter with HAWC Observations of dwarf Irregular galaxies}

\ShortTitle{HAWC DM searches on dIrr galaxies}

\author{\speaker{Sergio Hern\'andez Cadena}$^a$, Jos\'e Serna Franco$^b$, Rub\'en Alfaro Molina$^c$\\
        Instituto de F\'isica, Universidad Nacional Aut\'onoma de M\'exico\\
        E-mail: $^a$\email{shcdna@fisica.unam.mx}, $^b$\email{j\_serna@ciencias.unam.mx}, $^c$\email{ruben@fisica.unam.mx}}

\author{Viviana Gammaldi$^{1,2,4,5,d}$, Ekaterina Karukes$^{3,e}$, Paolo Salucci$^{1,f}$\\
        $^1$Scuola Internazionale Superiore di Studi Avanzati (SISSA), $^2$Instituto Nazionale di Fisica Nucleare (INFN), $^3$Astrocent, Nicolaus Copernicus Astronomical Center Polish Academy of Sciences, $^4$Departamento de F\'isica Te\'orica, Universidad Aut\'oonoma de Madrid, $^5$58 Instituto de F\'isica Te\'orica UAM-CSIC\\
        E-mail: $^d$\email{viviana.gammaldi@uam.es}, $^e$\email{ekarukes@camk.edu.pl}, $^f$\email{salucci@sissa.it}}
        
\author{for the HAWC Collaboration\\
        For a complete author list, see  https://www.hawc-observatory.org/collaboration/icrc2019.php\footnote{for collaboration list see PoS(ICRC2019)1177}\\
        }

\abstract{It has been shown that the dynamics of dwarf Irregular (dIrr) galaxies are dominated by dark matter. It is also observed that these galaxies have low star formation rates and metallicities, and no gamma-ray emission at ultra very high energies is expected. Because of their distance, dark matter content and vast number, dIrr galaxies are good targets to perform indirect dark matter searches by ground-based and wide field of view gamma-ray experiments, like HAWC. We analyzed data at the position of 31 dIrr galaxies within the HAWC field-of-view and no significant excess was found. Here, we present the individual and combined limits on the annihilation cross-section and decay lifetime of weakly interacting massive particles with mass between 1 and 100 TeV.}

\FullConference{36th International Cosmic Ray Conference -ICRC2019-\\
		July 24th - August 1st, 2019\\
		Madison, WI, U.S.A.}

\begin{document}

\section{Introduction}

While some indirect dark matter searches are focused on observing a small number of targets, wide field of view experiments can perform these searches on multiple populations at the same time. In particular, a combined analysis leads to stronger constraints on physical parameters of dark matter candidates. At the TeV scale, one population of interest is dwarf irregular (dIrr) galaxies. Several studies showed that these galaxies are dominated by dark matter. It is then plausible to expect a positive signature of gamma-rays from annihilation or decay of dark matter. Unlike dwarf spheroidal (dSph) galaxies, dIrr galaxies have gas and star-formation regions which can contribute to the gamma-ray flux. Because of the environmental characteristics in dIrr galaxies, it is expected that the gamma rays produced have energies in the GeV range \cite{martin,sfrhess}. Above 1 TeV this gamma-ray flux is negligible, and we treat dIrr galaxies as background-free targets. The present approach allows us to apply the analysis used for dSph galaxies to dIrr ones.\\
In this work, we used a sample of 31 dIrr galaxies within the HAWC field-of-view, see Table \ref{tabData}. We selected these galaxies from \cite{karukes}. As a result of analysis using kinematic data, dIrr galaxies have dark matter haloes described by a Burkert profile. We computed the astrophysical factors, both for annihilation and decay, using the Clumpy package \cite{clumpy}. After that, we obtained the gamma-ray flux using the photon spectra for channels $b\bar{b}$ and $\tau^{+}\tau^{-}$. The Pythia software \cite{pythia} is used to compute the photon spectra, as in previous HAWC analysis \cite{hawc_dsph}.

\begin{table}[ht!]
\begin{center}
\begin{tabular}{l||c|c|c|c}
Name & R.A. & DEC. & $\log_{10}(\frac{\mathrm{J}}{\text{TeV}^2\text{cm}^5})$ & $\log_{10}(\frac{\mathrm{D}}{\text{TeV}\text{cm}^2})$ \\
& (deg) & (deg) & & \\ 
\hspace{0.5cm}(1) & (2) & (3) & (4) & (5) \\
And IV &10.62 & 40.57 & 9.764 & 13.463 \\
DDO 101 & 177.91 & 31.51 & 10.356 & 14.312 \\
DDO 125 & 186.92 & 43.493 & 10.467 & 14.165\\
DDO 133	& 188.22 & 31.54 & 11.501 & 15.274 \\
DDO 154	& 193.52 & 27.15 & 11.800 & 15.397 \\
DDO 168 & 198.61 & 45.91 & 11.365 & 15.271 \\
DDO 43 & 112.07 & 40.77 & 10.109 & 13.853 \\
DDO 52 & 127.11 & 41.85 & 10.452 & 14.401 \\
Haro 29 & 186.56 & 48.49 & 9.974 & 13.764 \\
Haro 36 & 191.73 & 51.61 & 10.642 & 13.581 \\
IC 10 & 5.10 & 59.29 & 11.857 & 15.619 \\
IC 1613 & 16.19 & 2.13 & 11.632 & 15.325 \\
NGC 3741 & 174.02 & 45.28 & 9.814 & 13.417 \\
NGC 6822 & 296.23 & -14.80 & 12.173 & 15.943 \\
UGC 11583 & 307.56 & 60.44 & 10.676 & 14.605 \\
UGC 1281 & 27.38 & 32.59 & 10.854 & 14.739 \\
UGC 1501 & 30.31 & 28.84 & 10.937 & 14.843 \\
UGC 2455 & 194.92 & 25.23 & 10.392 & 14.250 \\
UGC 5272 & 147.59 & 31.48 & 10.721 & 14.731 \\
UGC 5427 & 151.17 & 29.36 & 10.133 & 14.007 \\
UGC 5918 & 162.40 & 65.53 & 10.512 & 14.420 \\
UGC 7047 & 181.01 & 52.58 & 10.630 & 14.444 \\
UGC 7232 & 183.43 & 36.63 & 10.845 & 14.581 \\
UGC 7559 & 186.77 & 37.14 & 11.105 & 14.938 \\
UGC 7603 & 187.18 & 22.82 & 11.368 & 15.251 \\
UGC 7861 & 190.46 & 41.27 & 10.804 & 14.715 \\
UGC 7866 & 190.56 & 38.50 & 10.672 & 14.462 \\
UGC 7916 & 191.10 & 34.38 & 11.012 & 14.898 \\
UGC 8508 & 202.68 & 54.91 & 10.362 & 14.071 \\
UGC 8837 & 208.68 & 53.90 & 10.856 & 14.802 \\
WLM & 0.49 & -15.46 & 12.062 & 15.777 \\
\end{tabular}
\caption{Sample of dIrr galaxies. We show the 31 dIrr galaxies within the HAWC field-of-view used in this study. Columns: name of the galaxy (1), the right ascension ($\alpha$) (2) and declination ($\delta$) (3) of the galaxy, the astrophysical factor for annihilation (4) and decay (5), computed with \textsc{Clumpy} \cite{clumpy}.}    
\label{tabData}
\end{center}
\end{table}

\section{Data and Analysis}

We analyzed HAWC data from 1017 transits. We divided the analysis into two classes: individual and combined analysis. For the individual analysis, a maximum Likelihood method is used for each individual dIrr galaxy to obtain the best fit parameters of the model including annihilation or decay of dark matter particles (signal model). The Likelihood is calculated using:
\begin{equation}
L_{\text{Signal}} = \prod_i\ln\left(\frac{1}{N_i!}(B_i+Si)^{N_i}\times e^{B_i+S_i}\right)
\end{equation}
where $S_i$ is the number of expected events produced by annihilation or decay of dark matter particles, $B_i$ is the number of background counts observed, and $N_i$ is the total number of events observed in a bin.\\
We also computed the Likelihood of the Signal model with respect to the Null hypothesis ($L_{\text{Null}}=L_{\text{Signal}}(S_i=0)$). We do not observe any significant excess at the position of all dIrr galaxies. Therefore we calculated the exclusion limits at 95\% C.L.\\
In the combined analysis, we used the joint-likelihood approach to compute the value of the parameter common to all signal models, either annihilation cross-section or decay lifetime. As in Individual analysis, we compute the significance of the joint model with respect to the null hypothesis. We did not observe any significant excess and the significance was converted into an exclusion limit.\\
We used the Liff package to compute the significance and exclusion limits. Liff is part of the HAWC internal software. For more details about the HAWC Observatory and the analysis see \cite{hawc_dsph,liff}.

\section{Results and Discussion}

We present the exclusion limits for the galaxies in our sample in Figures 1 (annihilation) and 2 (decay).  In both figures, the solid-black line shows the combined limit using the 31 galaxies. We can observe that galaxy DDO 154 gives the best limit because of its position in a declination band where HAWC has good sensitivity. Assuming a dark matter particle having mass of 10 TeV, $\langle\sigma_{\chi}v\rangle = 3.31\times10^{-22}\text{cm}^3\text{s}^{-1}$ for annihilation to $\tau^+\tau^-$ leptons, and $\tau_{\chi}=6.16\times10^{+25}\text{s}$ for decay to $\tau^+\tau^-$ leptons. We can observe that the combined limit shows two behaviors: for energies < 10 TeV, the combined limit is the same or even worse than the DDO 154 limit; for energies > 10 TeV, the combined limit is better than DDO 154 limit.The combined limit is not better than the DDO 154 limit below 10 TeV because of the large dispersion in the values of the signal-to-background ratio of dIrr galaxies.

\begin{figure}[ht]
    \centering
    \includegraphics[width=0.49\linewidth]{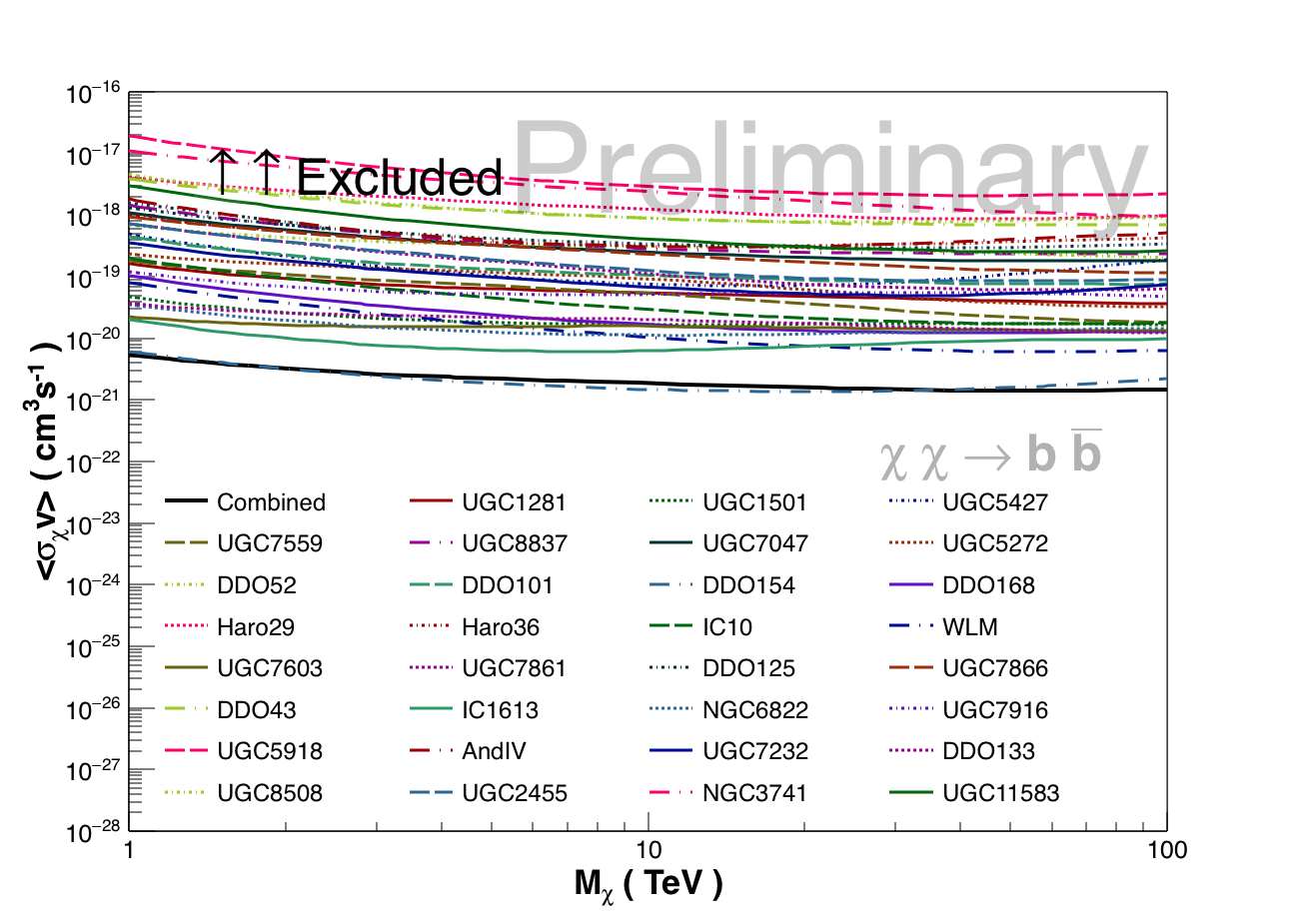}
    \includegraphics[width=0.49\linewidth]{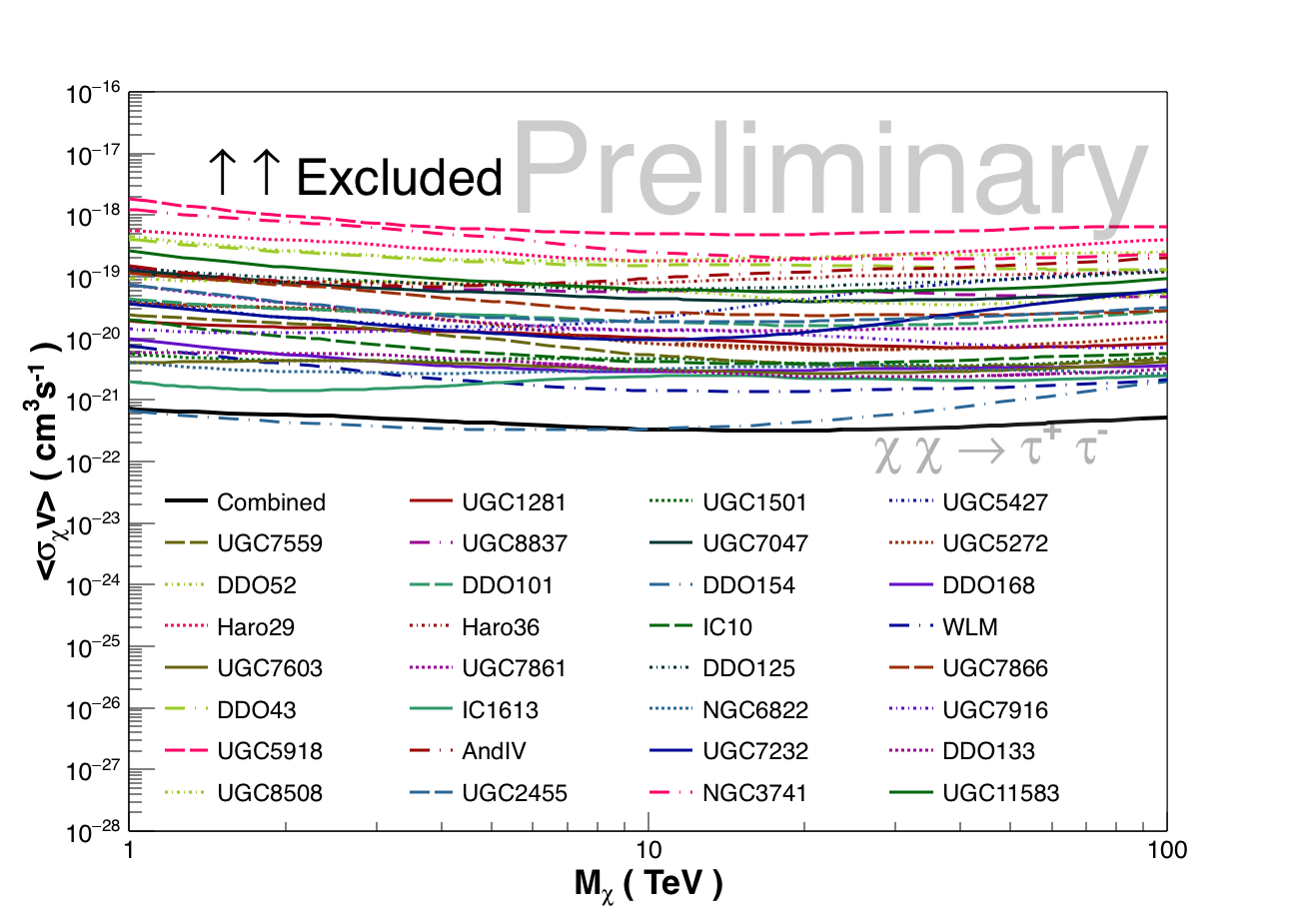}
    \caption{Exclusion limits for annihilation of dark matter particles. We show the individual limits for 31 dIrr galaxies and the combined limit (solid black line) for two different channels. Left: Annihilation to $b\bar{b}$ quarks. Right: Annihilation to $\tau^{+}\tau^{-}$ leptons.}
    \label{anaIndLimits}
\end{figure}

\begin{figure}[ht]
    \centering
    \includegraphics[width=0.49\linewidth]{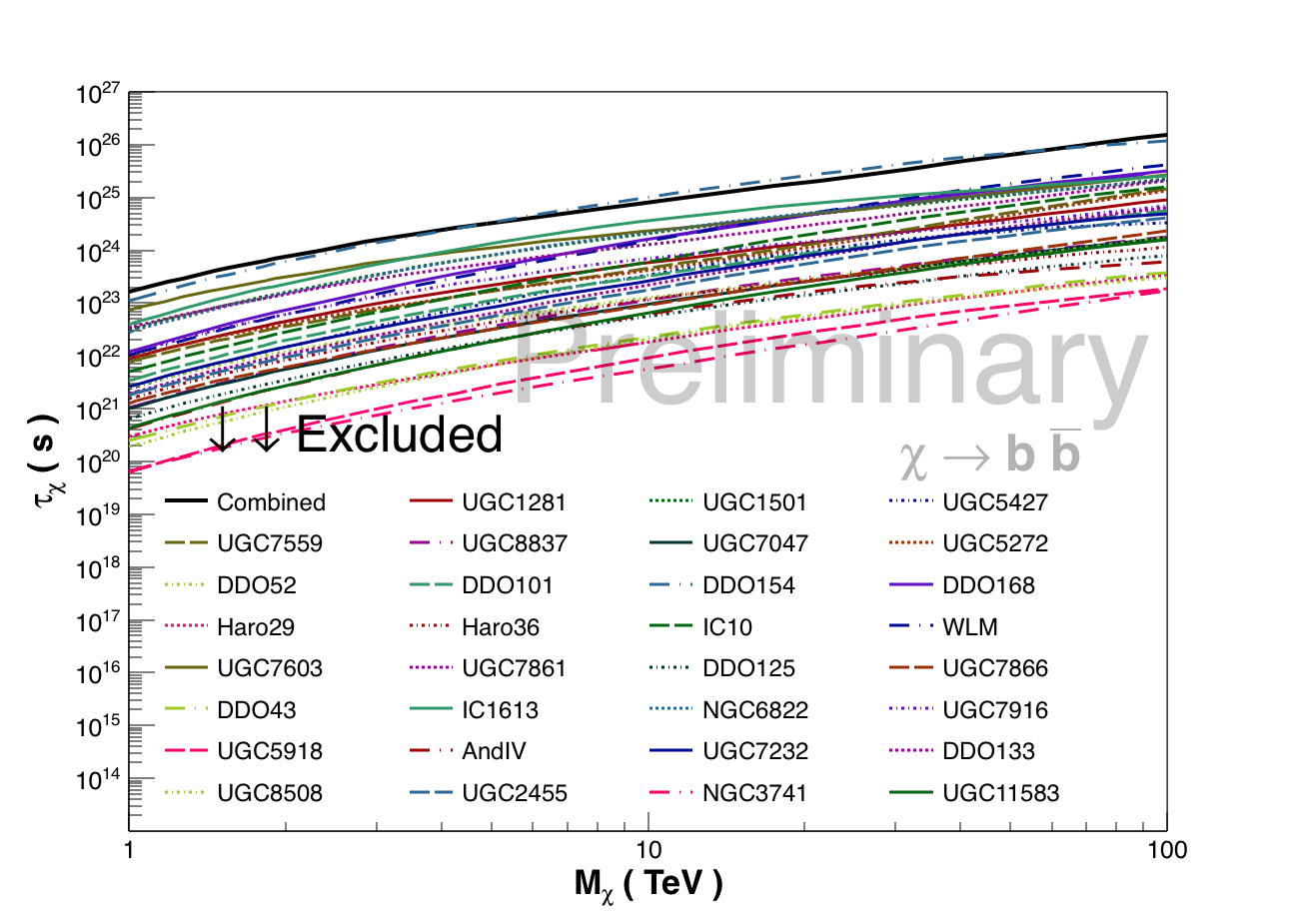}
    \includegraphics[width=0.49\linewidth]{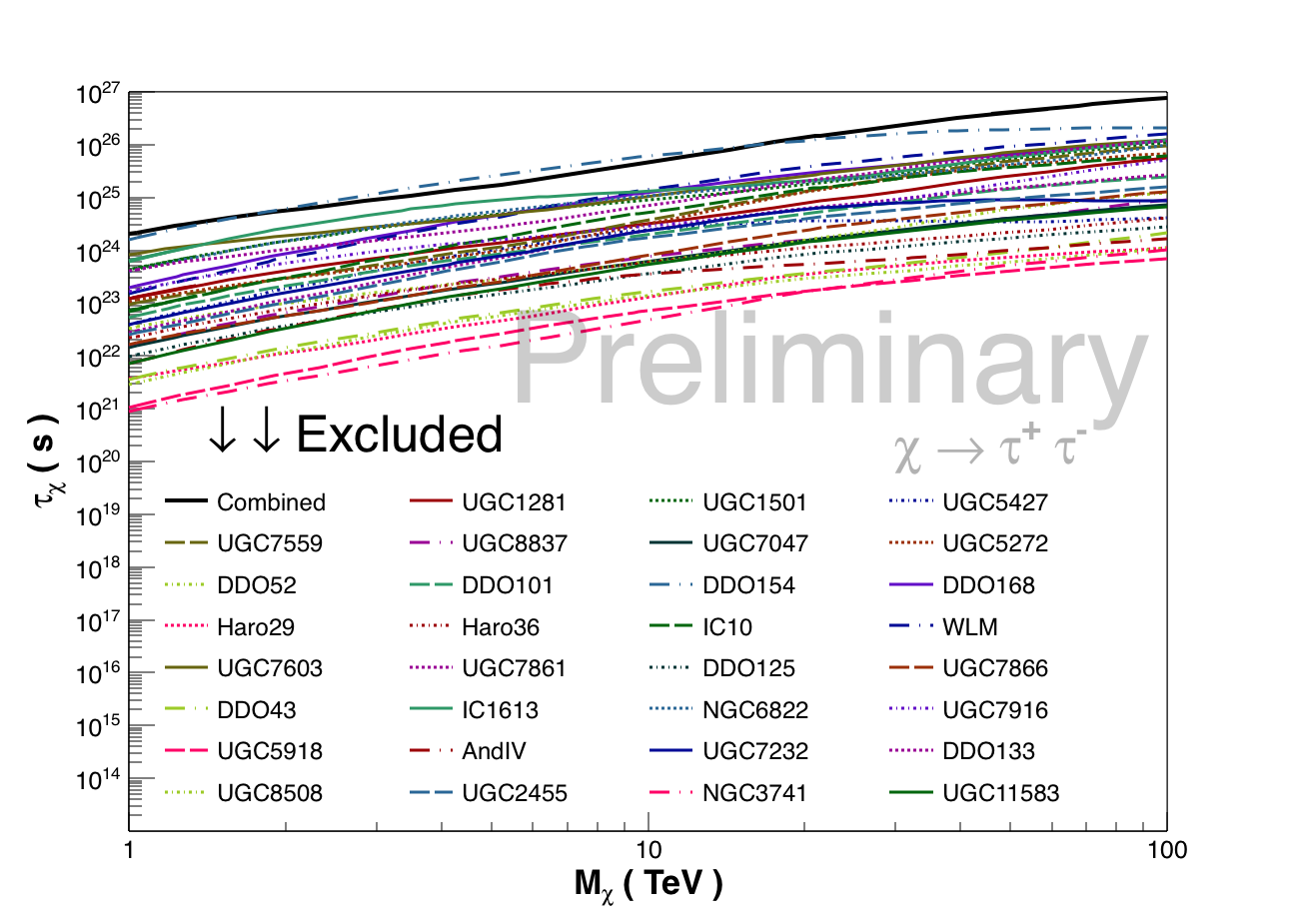}
    \caption{Exclusion limits for decay of dark matter particles. We show the individual limits for 31 dIrr galaxies and the combined limit (solid black line) for two different channels. Left: Decay to $b\bar{b}$ quarks. Right: Decay to $\tau^{+}\tau^{-}$ leptons.}
    \label{decIndLimits}
\end{figure}

\begin{figure}[ht]
    \centering
    \includegraphics[width=0.49\linewidth]{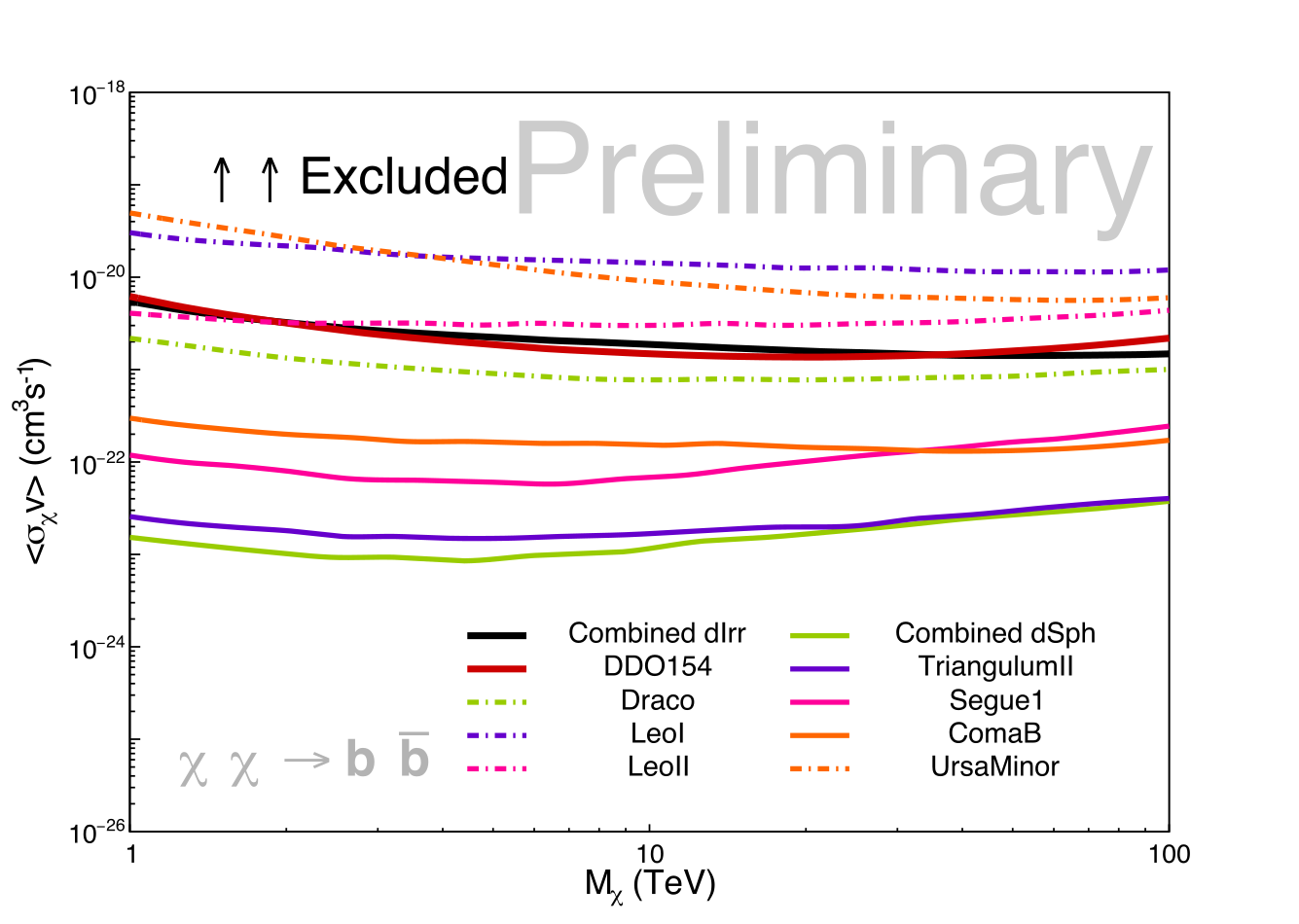}
    \includegraphics[width=0.49\linewidth]{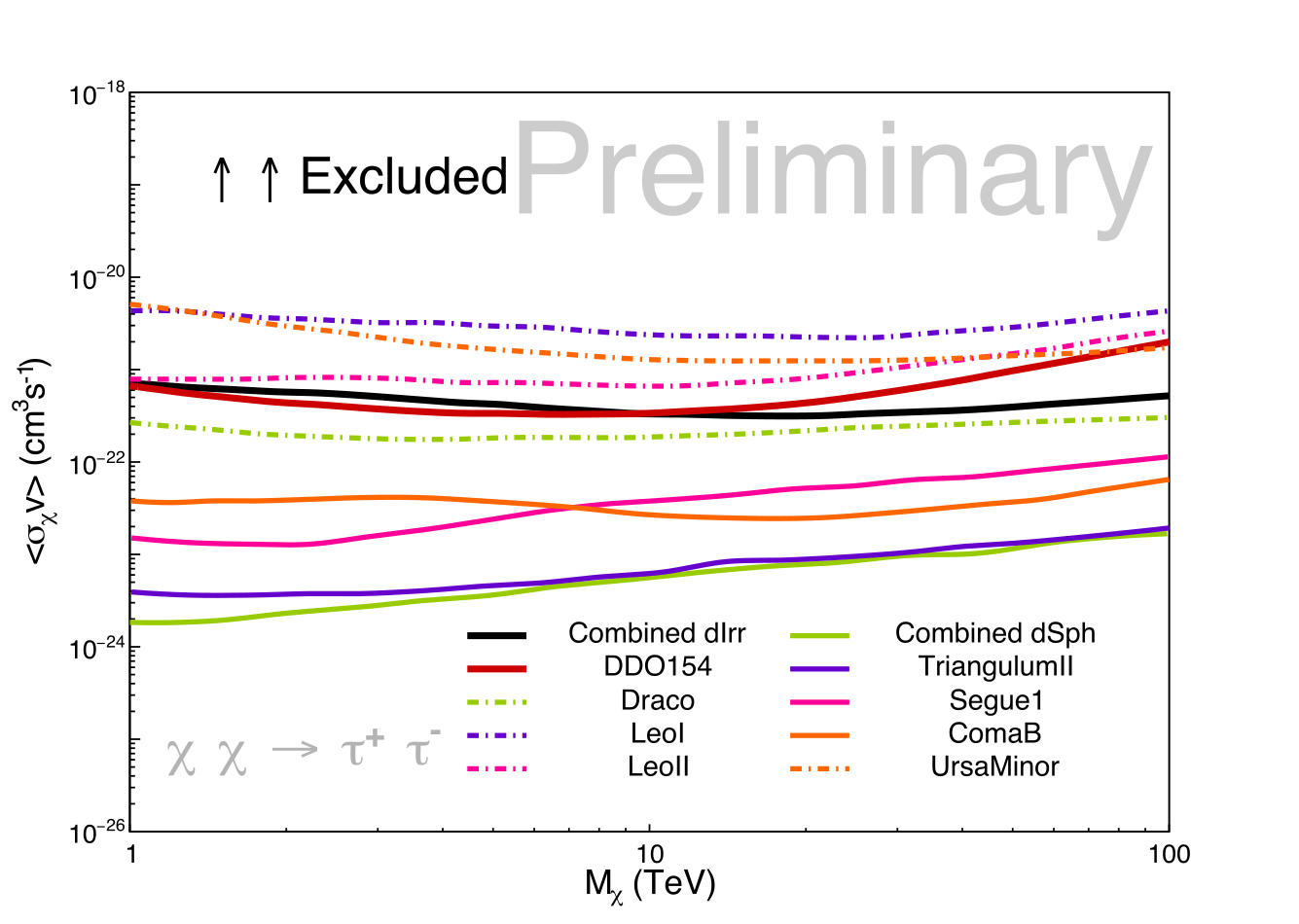}
    \caption{Exclusion limits for annihilation of dark matter particles. We show the individual limit for DDO 154 galaxy and the combined limit (solid black line) for two different channels. Left: Annihilation to b quarks. Right: Annihilation to $\tau$ leptons. We also show the limits for 7 dSph galaxies previously analyzed by HAWC.}
    \label{anaComp}
\end{figure}

\begin{figure}[ht]
    \centering
    \includegraphics[width=0.49\linewidth]{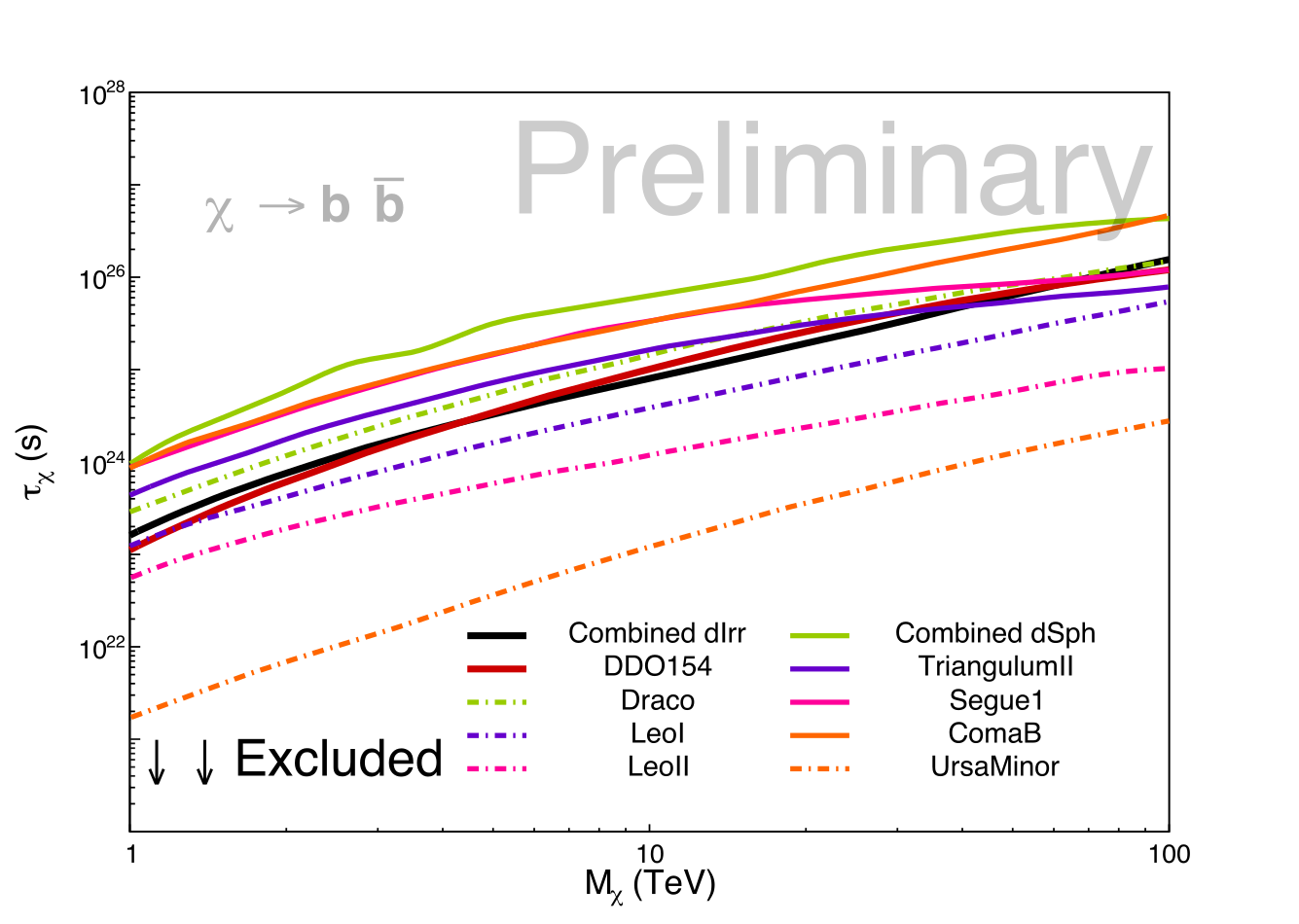}
    \includegraphics[width=0.49\linewidth]{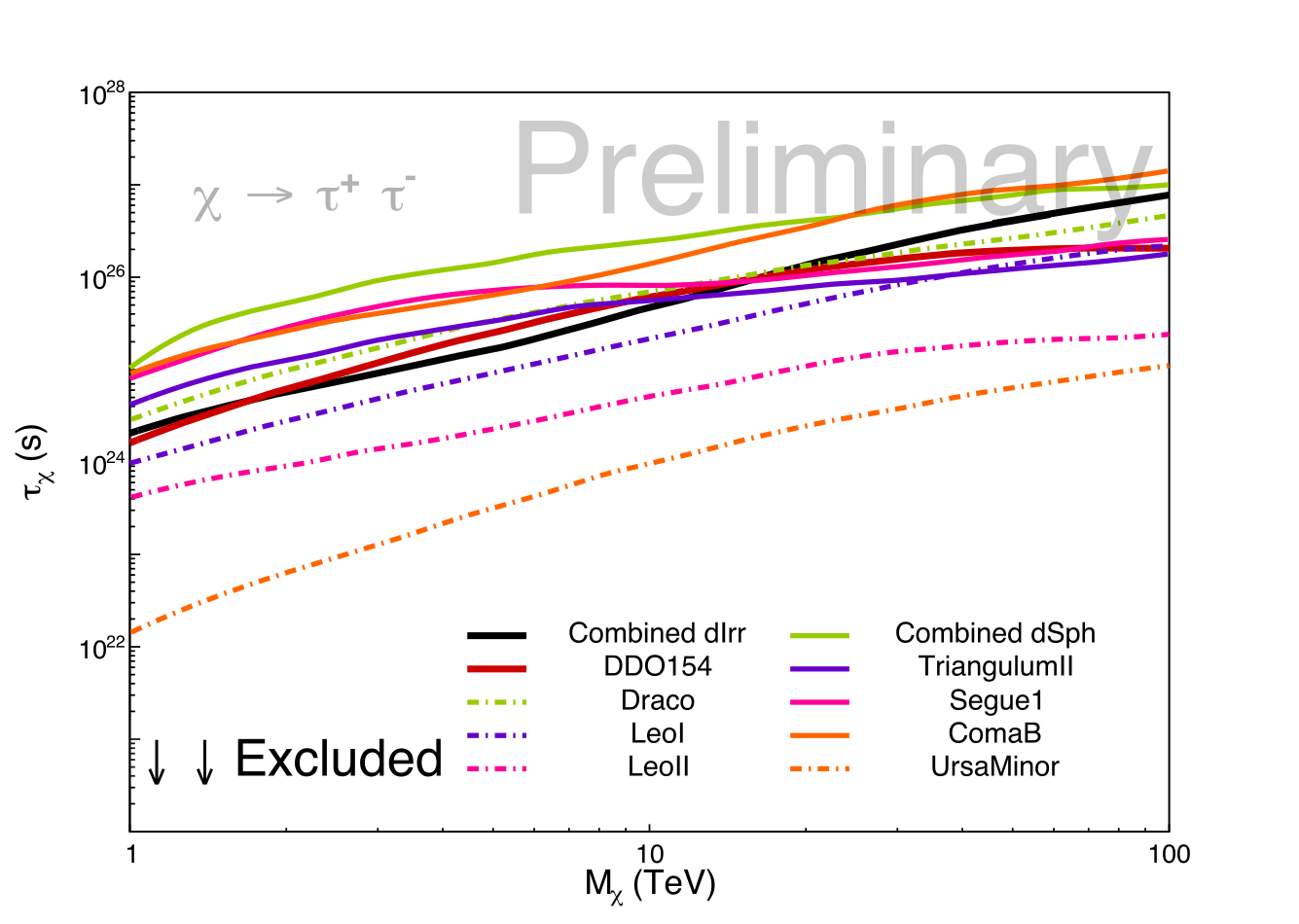}
    \caption{Exclusion limits for decay of dark matter particles. We show the individual limit for DDO 154 galaxy and the combined limit (solid black line) for two different channels. Left: Annihilation to b quarks. Right: Annihilation to $\tau$ leptons. We also show the limits for 7 dSph galaxies previously analyzed by HAWC.}
    \label{decComp}
\end{figure}

In Figures \ref{anaComp} and \ref{decComp}, we show a comparison between the HAWC dIrr and dSph exclusion limits. We present the exclusion limits for four classical dSph galaxies (dashed lines) and three ultrafaint dSph galaxies (solid lines). We also show the combined limits obtained for 15 dSph galaxies within the HAWC filed-of-view. For annihilation of dark matter (Figure \ref{anaComp}), we observe that the combined limit for dIrr galaxies (solid black line) and the DDO 154 limit (solid red line) are comparable to the exclusion limits obtained for classical dSphs, but not to limits for ultra-faint galaxies which have large values of J-factor in comparison to dIrr galaxies. For decay (Figure \ref{decComp}), the exclusion limits for dIrrs are comparable to both populations, classical and ultrafaint galaxies. For decaying dark matter,  it is possible to perform a combined analysis using a sample of dIrr and dSph galaxies to constrain the lifetime of dark matter candidates with masses at TeV scale.

\section{Conclusions}

We show that dIrr galaxies comprise a suitable population to perform indirect dark matter searches.This analysis does not include detailed kinematics to compute J-factors and is therefore limited. The actual number of dIrr galaxies within the local volume is four times bigger than the sample of 30 we used here. Including more dIrr galaxies in the combined analysis will improve the constraints on annihilation cross-section and decay lifetime of dark matter candidates. The combined analysis presented here is possible because of the characteristics of the HAWC Observatory, which can observe a large portion of the sky and test dark matter models in several populations simultaneously.


\section{Acknowlegments}

We acknowledge the support from: the US National Science Foundation (NSF) the US Department of Energy Office of High-Energy Physics; the Laboratory Directed Research and Development (LDRD) program of Los Alamos National Laboratory; Consejo Nacional de Ciencia y Tecnolog\'{\i}a (CONACyT), M{\'e}xico (grants 271051, 232656, 260378, 179588, 239762, 254964, 271737, 258865, 243290, 132197, 281653)(C{\'a}tedras 873, 1563, 341), Laboratorio Nacional HAWC de rayos gamma; L'OREAL Fellowship for Women in Science 2014; Red HAWC, M{\'e}xico; DGAPA-UNAM (grants AG100317, IN111315, IN111716-3, IA102715, IN111419, IA102019, IN112218); VIEP-BUAP; PIFI 2012, 2013, PROFOCIE 2014, 2015; the University of Wisconsin Alumni Research Foundation; the Institute of Geophysics, Planetary Physics, and Signatures at Los Alamos National Laboratory; Coordinaci{\'o}n de la Investigaci{\'o}n Cient\'{\i}fica de la Universidad Michoacana; Polish Science Centre grant DEC-2014/13/B/ST9/945, DEC-2017/27/B/ST9/02272; Royal Society - Newton Advanced Fellowship 180385. Thanks to Scott Delay, Luciano D\'{\i}az and Eduardo Murrieta for technical support.

\end{document}